\documentclass[aps,prd,amssymb,eqsecnum,showpacs] {revtex4}

\begin{document}

\title{Global aspects of radiation memory}

\author{ J. Winicour${}^{1,2}$
       }
\affiliation{
${}^{1}$ Department of Physics and Astronomy \\
        University of Pittsburgh, Pittsburgh, PA 15260, USA\\
${}^{2}$ Max-Planck-Institut f\" ur
         Gravitationsphysik, Albert-Einstein-Institut, \\
	 14476 Golm, Germany \\
	 }

\begin{abstract}

Gravitational radiation has a memory effect represented by a net change
in the relative positions of test particles.  Both the linear and nonlinear sources
proposed for this radiation memory are of the  "electric" type, or E mode, as
characterized by the even parity of the polarization pattern. Although "magnetic" type,
or B mode, radiation memory is mathematically
possible, no physically realistic source has been identified. There is an
electromagnetic counterpart to radiation memory in which the velocity
of charged test particles obtain a net "kick". Again, the physically realistic sources of
electromagnetic radiation memory that have been identified are of the electric
type. In this paper, a global null cone description of  the electromagnetic field 
is applied to establish the non-existence of B mode radiation memory and the
non-existence of E mode radiation memory due to a bound charge distribution.

\end{abstract}


\maketitle

\section{Introduction}

The memory effect due to radiation fields first received attention in the case of
gravitational radiation, where it was noted that certain radiating systems would
produce a net displacement between test particles in the wave zone~\cite{zeld}. The
effect was initially recognized in linearized gravity where the radiation memory arises
from an exploding system of massive particles which escape to infinity. However,
in the fully nonlinear gravitational  case, another source of radiation memory was
noted by Christodoulou~\cite{christo}, which was interpreted as arising from the flux
of radiation energy to infinity instead of particles.~\cite{thorn,will} In order to clarify
the relation between these linear and nonlinear sources of gravitational radiation
memory, the analogous effect has been studied in the case of electromagnetic
radiation~\cite{bieri1}, and even scalar radiation~\cite{wald},
where electromagnetic radiation memory gives charged test particles
a net kick, i.e. a net momentum recoil. There is a global feature of radiation
memory that has not yet been explored. All the known examples produce
radiation of the ``electric'' type, as distinguished by the even parity of the
polarization pattern. Here we apply a null cone formulation of Maxwell's equations as a
characteristic initial value problem for the vector potential to show that electromagnetic
radiation memory of the  odd parity ``magnetic'' type cannot be produced by a physically
realistic charge-current distribution.

Our approach is based upon a treatment of the coupled Einstein-Maxwell equations~\cite{thesis}
which generalizes the null hypersurface formulation of the Einstein equations pioneered
by Bondi et al~\cite{bondi} and by Sachs~\cite{sachs}.
The corresponding treatment of the Maxwell field in the flat space case as a characteristic
initial value problem in a null gauge~\cite{tam} is reviewed in Sec.~\ref{sec:maxwell}.
This approach is useful for investigating the sky pattern of the electromagnetic radiation.
In this respect, it has an advantage over the traditional Green function approach where the
radiation field in a given direction is determined by a retarded time integral over the 
corresponding null hyperplane; i.e. the radiation field of a scalar field $\Phi$
in the direction of  the unit vector ${\mathbf n}$ produced by a source $S$ is
\begin{equation}
             r\Phi(t,r,{\mathbf n}) =\int S(t-r+{\mathbf n}\cdot {\mathbf x'}, {\mathbf x}')dV' , 
             \quad r\rightarrow \infty ,
\end{equation}
where the integral is over a null hyperplane in the ${\mathbf n}$ direction.
In order to reveal the full sky pattern of the radiation at a given time, it is necessary to convert
the retarded time  dependence on the null-hyperplane to ordinary time by, say, restriction of
the source to a point particle, or by a Fourier decomposition of the time dependence of the
source or by a slow-motion approximation. Except for the point particle case, these methods
are not well-adapted to studying the sky pattern of the long  term memory
effect.

We follow the common electromagnetic nomenclature and refer to
the even and odd parity of the radiation
patterns as  E mode (``electric'' type) or B mode
(``magnetic'' type), respectively. Of course, in the electromagnetic case, the electric and magnetic
radiation fields are equal in magnitude and orthogonal in direction. The 
E or B modes refer to the parity of the sky pattern of the polarization.
This distinction was the crucial element in the recent BICEP2 experiment~\cite{bicep2},
in which a B mode component detected in the cosmic microwave background is possible
evidence of the existence of primordial gravitational waves.
  
The distinction between E and B mode radiation memory
is  elaborated in more technical detail in Sec.~\ref{sec:emrm},
where examples of electromagnetic  memory of both types are presented. These examples show 
that B mode memory can be generated by ingoing radiation fields. 
However, all the proposed charge-current sources of electromagnetic 
radiation memory only  produce E mode memory. The same is true for examples
of gravitational memory, as discussed in Appendix A. 
This raises the question whether B mode memory must be of primordial origin or generated by
ingoing waves produced in the distant past.
 
We investigate these issues in the context of the null gauge formulation of Maxwell equations
in Sec.~\ref{sec:em} for E mode radiation and in Sec.~\ref{sec:bm} for B mode radiation.
Our chief results are:
\begin{itemize}

\item E mode memory cannot be generated by a physically
realistic bound charge and current distribution, i.e. a spatially bounded distribution
of charges and currents whose electric and magnetic fields approach stationary
values at infinite past and future retarded times. The generic physically
realistic source of E mode radiation memory is a system
of charged particles with escape velocity.

\item For B mode radiation memory, no such physically realistic sources exist.

\end{itemize}
The stationarity condition is necessary to eliminate radiation memory due to
source free waves.

\section{Maxwell's equations in the null gauge}
\label{sec:maxwell}

It will be useful to refer to three separate coordinate systems. A Cartesian inertial system
$(t,x^i)=(t,x,y,z)$, the associated spherical coordinate system $(t,r,x^A)=(t,r,\theta,\phi)$,
$r^2 =x^2+y^2+z^2$,  and the associated outgoing null coordinate system
$x^\alpha=(u,r,x^A)$, with retarded time $u=t-r$ and vertices of the null cones at $r=0$.
In these retarded null coordinates, the Minkowski metric is
\begin{equation}
     ds^2= -du^2 -2du dr + r^2 q_{AB} dx^A dx^B,
\end{equation}
with contravariant components     
\begin{equation}
      g^{ur}=-1, \quad g^{rr} =1, \quad g^{AB} =r^{-2} q^{AB}, \quad  g^{uu}= g^{uA}=0,
 \end{equation}
where $q_{AB}dx^A dx^B =d \theta^2 +\sin^2 \theta d\phi^2$ is the unit sphere metric. 

Referred to these coordinates, Maxwell equations
\begin{equation}
     {\cal E}^\alpha:= \nabla_\beta F^{\alpha\beta} -4\pi J^\alpha =0
\end{equation}
consist of the three {\it main equations}
\begin{equation}
    {\cal E}^u= 0, \quad  {\cal E}^A= 0
\end{equation}
and the {\it supplementary condition} ${\cal E}^r= 0$.
If the main equations are satisfied then charge-current conservation 
$\nabla_\alpha J^\alpha=0$ implies 
\begin{equation}
    \partial_r (r^2 {\cal E}^r)= 0.
\end{equation}
As a result, the supplementary condition is satisfied if it is satisfied on any $r=const$ sphere.
In particular, it is satisfied if fields at the vertices $r=0$ are non-singular.

We treat Maxwell's equations in terms of a vector potential
\begin{equation}
     F^{\alpha\beta} =\nabla^\alpha A^\beta -\nabla^\beta A^\alpha
\end{equation}
and use the gauge freedom $A_\alpha \rightarrow A_\alpha+\partial_\alpha \Lambda$
to introduce a null gauge
\begin{equation}
   A_r =A^u =0.
\end{equation}
This can be arranged by the gauge choice
\begin{equation}
    \Lambda (u,r,x^A) = -\int_0^r A_r dr.
    \label{eq:nrgauge}
\end{equation}
In a similar way, we could pin down the remaining gauge freedom  $\Lambda (u,x^A)$
to set
\begin{equation}
   A_u(u,r,x^A)|_{r=0} = 0,
   \label{eq:nugauge}
\end{equation}
which is important for posing a unique characteristic evolution problem,
although this is not essential for the purpose of this paper.

Regularity conditions follow from the requirement
that the Maxwell field $F_{\alpha\beta}$ have smooth components
in Cartesian inertial coordinates. The transformation $(t,x^i) \rightarrow (u,r,x^A)$
then induces vertex regularity conditions in the retarded null spherical coordinates.
The coordinate derivatives are related by
\begin{equation}
  \partial_u=\partial_t \, , \quad \partial_r=\partial_t+n^i\partial_i \, ,
  \quad  \partial_A=r (\partial_A n^i)\partial_i \, ,
\end{equation}
where
\begin{equation}
     n^i=\frac{x^i}{r} = (\sin\theta\cos\phi,\sin\theta\sin\phi,\cos\theta)
\end{equation} 
is composed of $\ell=1$ spherical harmonics.
The components of  $F_{\alpha\beta}$ transform according to
 \begin{eqnarray}
     F_{ru} &=& \partial_r A_u = n^i F_{it} 
     \label{eq:rureg}\\
      F_{rB} &=& \partial_r A_B = r (\partial_B n^j)( F_{tj} + n^i F_{ij})
        \label{eq:rBreg} \\
       F_{Bu} &=& \partial_B A_u -\partial_u A_B = r (\partial_Bn^i )F_{it} 
         \label{eq:Bureg}\\
        F_{BC} &=& \partial_B A_C-  \partial_C A_B= r^2  (\partial_Bn^i ) (\partial_C n^j ) F_{ij} .
          \label{eq:BCreg}
 \end{eqnarray}
Regularity at  the vertex of the null cone requires $ \partial_r A_B |_{r=0}=0$, 
$( \partial_B A_u-\partial_u A_B)|_{r=0}=0$ and $(\partial_B A_C-  \partial_C A_B)|_{r=0}=0$.

At future null infinity ${\mathcal I}^{+}$, the
asymptotic behavior of the Maxwell field for an isolated system implies
$F_{ru} =\partial_r A_u= O(r^{-2})$,  $F_{Bu}=\partial_B A_u-\partial_u A_B = O(1)$ and
$F_{rB}=\partial_r A_B= O(r^{-1})$. Note, however, that in the null
gauge the Coulomb field of a moving charge has the asymptotic dependence
\begin{equation}
       A_B \sim a_B (x^C) \ln r.
\end{equation} 

The angular components of the electric field are $E_B =F_{Bu} $. In these spherical
coordinates $E_B=O(1)$ at null infinity so that the net momentum ``kick'' on a test
particle with charge $q$ is determined by
\begin{equation}
    q  \int_{-\infty}^{\infty} E_B du = q  \int_{-\infty}^{\infty} \partial_B A_u du -q \Delta A_B,
\end{equation}
where we use the notation  $\Delta F(x^C)=F(u,x^C)|_{u=\infty}-F(u,x^C)|_{u=-\infty}$.
Note that in terms of a orthonormal basis this kick
falls off as $O(1/r)$.

Because of the extraneous $r$-dependence in the angular components of
the metric in spherical coordinates, it is convenient to raise and
lower angular indices with the unit sphere metric $q_{AB}$. In doing so, we adopt
the convention that $J^A$ and $A_B$ are the true tensorial components induced
by the transformation from Cartesian coordinates, e.g $A^B =q^{BC}A_C$.
For  notational simplicity, we also denote covariant derivatives with respect to the
unit sphere metric by a ``colon'', e.g.
\begin{equation}
    { A_B}^{:B} = \frac{1}{\sqrt{q}} \partial_C( \sqrt{q}  q^{BC} A_B), \quad q=\det(q_{AB}).
\end{equation}

With these conventions, the main equations take the form of  a hypersurface equation
 ${\cal E}^u=0$, 
\begin{equation}
     4\pi J^u= \frac{1}{r^2} \partial_r \{r^2 \partial_r A_u   -{A_B}^{:B} \} ,
     \label{eq:hyp}
\end{equation}
and a dynamical equation  ${\cal E}^B=0$,
\begin{equation}
     4\pi J^B= \frac{q^{BC}}{r^2} \partial_r \{ 2\partial_u A_C  - \partial_C A_u
                    -  \partial_r A_C\}
         + \frac{q^{BC}}{r^4}( A_{D:C}- A_{C:D})^{:D} .
         \label{eq:dyn}
\end{equation}

The supplementary condition ${\cal E}^r=0$, which is automatically satisfied 
by virtue of the main equations and the vertex
regularity conditions, takes the form
\begin{equation}
   r^2 {\cal E}^r = -  4\pi r^2  J^r +  \partial_r {A_B}^{:B} - ( \partial_u A_B -\partial_B A_u)^{:B}
                          -r^2 \partial_u \partial_r A_u=0.
                           \label{eq:supp}
\end{equation}

The main equations determine the following natural evolution scheme for a characteristic initial
value problem~\cite{tam}. Given the initial values of the angular components $A_B(u=0,r,x^C)$, the
hypersurface equation can be integrated radially to determine $A_u(u=0,r,x^C)$. In turn,
the dynamical equation can be integrated radially to determine  $\partial_u A_B(u=0,r,x^C)$. By means
of a finite difference approximation, this determines $A_B(\delta u,r,x^C)$. This procedure  can then be
iterated to produce a solution provided the finite difference approximation converges.
Such a convergent finite difference evolution algorithm has been successful for the analogous
characteristic initial value problem for the Einstein equations~\cite{winrev}.
The coupled Maxwell-Einstein
equations have this same hierarchical structure as a set of hypersurface and dynamical equations
which can be integrated sequentially in the radial direction.~\cite{thesis}

\section{E and B mode radiation memory}
\label{sec:emrm}

A vector field on the sphere $V_B$ may be expressed as the a sum of a gradient
and a curl,
\begin{equation}
              V_A = \Phi_{:A} + \Psi^{:A} \epsilon_{BA},
              \label{eq:decomp}
\end{equation}
where $\epsilon_{AB}=-\epsilon_{BA}$ is the alternating tensor. In other terminology, $\Phi$
and $\Psi$ correspond to the electric and magnetic parts of $V_A$, i.e. the E mode
with positive parity and the B mode with negative parity. Alternatively,
the decomposition can be made in terms of a spin-weight-1 function $V=q^A V_A$
by introducing a complex dyad $q^A$ according to
\begin{equation}
      q^{AB} = q^{(A}\bar q^{B)}, \quad  \epsilon_{AB} = i q_{[A}\bar q_{B]} , 
      \quad   \epsilon^{AB} = i q^{[A}\bar q^{B]} 
        \quad  q^A\bar q_A =2,
 \end{equation}
where in standard spherical coordinates, $\epsilon_{\theta\phi}=\sin\theta$.
Then (\ref{eq:decomp})  takes the form 
\begin{equation}
       V= q^A (\Phi_{:A} +i\Psi_{:A}) = \eth v,  \quad v =\Phi+i\Psi,
\end{equation}
where $\eth $ is the spin-weight raising operator~\cite{goldberg}. Thus $\Phi$ is the real part
and $\Psi$ is  the  imaginary part of the spin-weight-0 potential $v$ for the
spin-weight-1 function $V$.

In order to treat the vector potential describing the exterior
radiation field emitted by an isolated system we introduce a Hertz
potential with the symmetry
\begin{displaymath}
        H^{\alpha\beta}=H^{[\alpha\beta]}.
\end{displaymath}
Then the vector potential
\begin{displaymath}
     A^\alpha=\partial_\beta H^{\alpha\beta}
\end{displaymath}
satisfies the Lorentz gauge condition and generates a solution of
Maxwell's equations provided the Hertz potential satisfies the wave
equation. 

Any source free wave can be generated in this way. Here 
we concentrate on dipole waves oriented with respect to the $z$-axis. Let
$(T^\alpha, X^\alpha, Y^\alpha, Z^\alpha)$ be a normalized basis aligned
with the axes of a Cartesian inertial frame.
The choice 
\begin{equation}
 H^{\alpha\beta}=(T^\alpha Z^\beta-Z^\alpha T^\beta)\frac{f(t-r)}{r}
 \label{eq:ehertz}
 \end{equation}
gives rise to an outgoing  E mode dipole wave with vector potential
\begin{equation}
    A_\alpha=-\bigg(\frac{f'(t-r)}{r}+\frac{f(t-r)}{r^2}\bigg) T_\alpha \cos\theta
           -\frac{f'(t-r)}{r}Z_\alpha,
           \label{eq:hertze}
\end{equation}
where $f'(u)=df(u)/du$.
The components in null spherical coordinates in the null gauge are 
\begin{equation}
       A_u=  (\frac {2 f'(u)}{r} + \frac {f(u)}{r^2})\cos\theta ,
\end{equation}
\begin{equation}
       A_r=0 ,
\end{equation}
\begin{equation}
       A_B=-(f'(u)-\frac{f(u)}{r})(\cos \theta)_{:B},
\end{equation}
with
\begin{equation}
       E_B = F_{Bu}= \bigg(f''(u)+\frac{2f'(u)}{r}+\frac{f(u)}{r^2}\bigg )(\cos \theta)_{:B}.
\end{equation}
The choice
\begin{equation}
    H^{\alpha\beta}=(X^\alpha Y^\beta-Y^\alpha X^\beta)\frac{f(t-r)}{r},
\end{equation}    
corresponding to the 4-dimensional dual of (\ref{eq:ehertz}),  gives
rise to an outgoing B mode dipole wave with vector potential 
\begin{equation}
    A_\alpha=-\bigg(\frac{f'(t-r)}{r}+\frac{f(t-r)}{r^2}\bigg) 
    \bigg(\frac{yX_\alpha}{r}-\frac{xY_\alpha}{r}\bigg) .
     \label{eq:hertzm}
\end{equation}
The components in null spherical coordinates in the null gauge are 
\begin{equation}
       A_u= A_r=0 ,
\end{equation}
\begin{equation}
       A_B=-(f'(u)+\frac{f(u)}{r})(\cos \theta)^{:A} \epsilon_{AB}.
\end{equation}

In both the case of E mode and B mode dipole waves, the memory is non-zero
provided
$$\Delta f' = f'(u=\infty) -f'(u=-\infty)\ne0.
$$
The electric and magnetic
radiation fields are proportional to $f''(u)$, so a non-zero memory is consistent
with finite energy flux to infinity provided $f(u)$ has the asymptotic behavior
$f'(u) =O(1)$ and $f''(u)=O(u^{-1})$ for large $|u|$.

The source free outgoing radiation solutions (\ref{eq:hertze}) and (\ref{eq:hertzm})
are singular at $r=0$ but the singularity can be removed by a superposition of
outgoing and ingoing waves
\begin{equation}
      \frac{f(t-r)-f(t+r)}{r} = \frac{f(u)-f(v)}{r}, \quad v=t+r.
      \label{eq:sup}
\end{equation}            
For a smooth function $f$ such solutions are smooth at $r=0$.
This superposition also eliminates a problem associated with
secular growth at early or late times which would otherwise result
from the asymptotic  dependence $f(u) \sim u$, which is
necessary for a non-zero radiation memory.  However, anomalous
time dependence still results.
This can be seen from the representative example
\begin{eqnarray}
        f(u) &=&Au , \quad u \le -T \nonumber \\
        f(u) &=& F(u), \quad -T<u < T,  \\
        f(u) &=& Bu , \quad  u\ge  T, \nonumber
\end{eqnarray}
where $F(u)$ is chosen to produce a smooth radiative waveform.
At  early and late times, 
\begin{eqnarray}
        \frac {f(u)-f(v)} {r} &=& \frac{Au -Av}{r}= -2A , \quad u \le -T, \quad v\le-T  , \nonumber \\
       \frac {f(u)-f(v)} {r} &=& \frac{Bu -Bv}{r}= -2B , \quad u \ge T, \quad v\ge T ,  \nonumber
\end{eqnarray}
but in a neighborhood of spatial infinity
\begin{equation}
      \frac {f(u)-f(v)} {r} =\frac{Au -Bv}{r} =-2B +\frac{(A-B)u}{r}, \quad u \le -T, \quad  v\ge T.
\end{equation}
Thus anomalous time dependence occurs in a neighborhood of spatial infinity for $A\ne B$, 
i.e. for the case of non-zero memory. This time dependence carries over to the Maxwell field.
For the electric dipole wave (\ref{eq:hertze}) the superposition (\ref{eq:sup}) leads to
\begin{eqnarray}
  E_r=F_{ru} &=&-2 \bigg ( \frac{ f'(u)+f'(v)}{r^2} +\frac{f(u) -f(v)}{r^3} \bigg) \cos \theta
         \nonumber \\
         &=&-2 \bigg (\frac{ A-B}{r^2} +\frac{(A-B)u}{r^3} \bigg) \cos \theta ,
         \quad u \le -T, \quad  v\ge T.
\end{eqnarray}
Thus, the E mode memory produced by this source free wave
can be ruled out by requiring that the Maxwell field approach a stationary limit
as $u\rightarrow -\infty$ and also as $u \rightarrow +\infty$, which corresponds to the
limit $v \rightarrow +\infty$. The same conclusion holds true for the corresponding
B mode memory. Note that this asymptotic stationarity condition does not
rule out ingoing waves with vanishing radiation memory, e.g. waves for which
$f(v)$ has compact support.
 
In these source free examples, the radiation memory originates
from ingoing radiation from the infinite past.
The question whether electromagnetic radiation memory can be generated by a
physically realistic charge-current
distribution is addressed in the next two sections.

\section{E mode radiation memory}
\label{sec:em}

Now consider E mode radiation fields for which we can set $A_B = \partial_B \alpha$
and
$$
E_B= F_{Bu} =\partial_B A_u -\partial_u A_B = \partial_B (A_u - \partial_u \alpha).
$$
In particular $\partial_B A_C -\partial_C A_B=0$ so that
the main equations (\ref{eq:hyp}) and (\ref{eq:dyn}) then reduce to the hypersurface equation
\begin{equation}
     4\pi r^2 J^u=  \partial_r \{r^2 \partial_r A_u 
         -{A_B}^{:B} \}
         \label{eq:ehyp}
\end{equation}
and dynamical equation 
\begin{equation}
     4\pi r^2 J^B=q^{BC}\partial_r \{ 2\partial_u A_C  - \partial_C A_u
                    -  \partial_r A_C\}  .
                  \label{eq:edyn}
\end{equation}
The supplementary condition (\ref{eq:supp}) reduces to
\begin{equation}
      4\pi r^2  J^r =  \partial_r {A_B}^{:B} - ( \partial_u A_B -\partial_B A_u)^{:B}
                          -r^2 \partial_u \partial_r A_u .
\end{equation}

The chief equation governing the radiation memory results from combining
(\ref{eq:ehyp}) and (\ref{eq:edyn}) using charge-current conservation,
which takes the explicit form
\begin{equation}
           r^2\partial_u J^u + \partial_r (r^2 J^r) +r^2 {J^A}_{:A}=0.
\end{equation}
But this procedure leads exactly to the supplementary condition (\ref{eq:supp}).
Evaluated at ${\mathcal I}^+$, the asymptotic falloff of the Maxwell
field implies that the supplementary condition further reduces to
\begin{equation}
   \{ 4\pi r^2  J^r -{E_B}^{:B}  +r^2 \partial_u  \partial_r A_u \}|_{{\mathcal I}^+}=0.
\end{equation}
This is precisely the equation obtained by Bieri and Garfinkle~\cite{bieri1} by a purely
asymptotic approach based upon an expansion in $1/r$ (see also ~\cite{wald}).
As pointed out in~\cite{bieri1}, it shows that the radiation memory is governed by two sources,
\begin{equation}
    \int_{-\infty}^{\infty}{ E_B}^{:B} du |_{{\mathcal I}^+}
    = \Delta (r^2 \partial_r A_u) |_{{\mathcal I}^+} 
          + 4\pi \int_{-\infty}^{\infty}r^2  J^r  du |_{{\mathcal I}^+} .
    \label{eq:mem1}
\end{equation}

The first term on the right hand side of (\ref{eq:mem1}) represents the radiation memory due to
the change in the asymptotic Coulomb-type field. It corresponds to the ``linear''  memory in the
gravitational case. For  a particle with unit charge which is asymptotically at rest at $u=-\infty$ and is
ejected with non-zero final velocity $V$ at $u=\infty$, this term corresponds to a boosted
Coulomb field. For final velocity $V$ in the $z$-direction, the resulting Coulomb
field at $u=\infty$ in null spherical coordinates (and null gauge) has component
\begin{equation}
    F_{ru}=\partial_r A_u 
               = \frac{1}{R^{*2}}\bigg (\partial_r R^*  -(1-V\cos\theta)\partial_u R^*\bigg )
               =\frac{1-V^2}{R^{*3}}\bigg(r(1-V\cos\theta) - uV\cos\theta\bigg ) ,
\end{equation}
where
$$ R^{*2}= (1-V^2)(x^2+y^2) +(z-Vt)^2 = r^2(1-V\cos\theta)^2 +2uVr (V-\cos\theta) +V^2 u^2.
$$
At  ${\mathcal I}^+$,
\begin{equation}
   (r^2 F_{ru})|_{{\mathcal I}^+}  =\frac{1-V^2}{(1-V\cos\theta)^2 }.
\end{equation}
After subtracting out the initial Coulomb field,
\begin{equation}
 \int_{-\infty}^{\infty}{ E_B}^{:B} du |_{{\mathcal I}^+} 
        =\Delta (r^2 F_{ru})|_{{\mathcal I}^+}  =\frac{1-V^2}{(1-V\cos\theta)^2 } -1.
\end{equation}
The radiation memory is then given by
\begin{equation}
     \int_{-\infty}^{\infty}  E_B du |_{{\mathcal I}^+}    = \frac{-V}{1-V\cos\theta} \partial_B \cos\theta .
\end{equation}
In the slow motion approximation this has a pure dipole angular dependence.
Analogous results for a system of unbound charged particles can be obtained by superposition.

\subsection{Lack of E mode radiation memory from a bound system}

The above results follow from a purely asymptotic analysis at ${\mathcal I}^+$.
In order to address the question whether a bound system of charged particles can
produce radiation memory it is apparent from the examples in Sec.~\ref{sec:emrm} that memory
arising from ingoing
radiation from ${{\mathcal I}^-}$ must be eliminated. We enforce this by requiring that
the field strengths approach stationary values in the limits $u=\pm \infty$. We adapt the null coordinates
to the asymptotic stationarity.
As a result, all components of the Maxwell field must satisfy $\partial_u F_{\alpha\beta} =0$
in the limits $u=\pm \infty$. (Note that this asymptotic stationarity does not rule out the
Coulomb field of a moving charge.)
In terms of the vector potential, with $A_B=\partial_B \alpha$, this implies
\begin{eqnarray}
0&=&\partial_u ( \partial_u A_B -\partial_B A_u)|_{u=\pm \infty}
    =\partial_u  ( \partial_u \alpha - A_u)_{:B}|_{u=\pm \infty} , \\
0&=&\partial_u \partial_r A_B = \partial_u \partial_r \alpha_{:B}|_{u=\pm \infty}, \\
0&=&\partial_u \partial_r A_u|_{u=\pm \infty}.
\end{eqnarray}
For technical convenience, we assume that $\alpha$ has no $\ell = 0$ spherical harmonic
component, i.e.
\begin{equation}
     \alpha = \sum_{\ell=1, |m| \le \ell}^{\ell=\infty} \alpha_{\ell m} Y_{\ell m} .
\end{equation}
These conditions then imply the early and late time functional dependencies
\begin{eqnarray}
     (A_u - \partial_u \alpha) &\rightarrow &f_1(r,x^A),
     \label{eq:asym1} \\
       \partial_r \alpha  &\rightarrow & f_2(r,x^A), \quad  \partial_r \partial_u \alpha  \rightarrow 0,
       \label{eq:asym2} \\
     \partial_r A_u &\rightarrow & f_3(r,x^A).
     \label{eq:asym3}
\end{eqnarray}
In addition, the requirement of a bound system requires that the 4-current satisfies
\begin{equation}
     \partial_u J^u|_{u=\pm \infty} =   J^r|_{u=\pm \infty} ={J^A}_{:A}|_{u=\pm \infty} =0.
\end{equation}

As a result of these early and late time conditions, the hypersurface equation implies
\begin{equation}
     4\pi r^2 J^u |_{u=\pm \infty} = \partial_r (r^2 \partial_r A_u 
         -{\alpha_{:B}}^{:B} )|_{u=\pm \infty} ,
         \label{eq:asymh}
\end{equation}
the dynamical equation implies
\begin{equation}
     0=- \partial_r {(  A_u +\partial_r \alpha)_{:B}}^{:B}|_{u=\pm \infty} 
                      \label{eq:asymd}
\end{equation}
and the supplementary condition implies
\begin{equation}
      0  =  {( \partial_r \alpha -  \partial_u \alpha  + A_u)_{B}}^{:B}|_{u=\pm \infty} .
                         \label{eq:asyms}
\end{equation}

By combining (\ref{eq:asym1}) and  (\ref{eq:asym2}) with (\ref{eq:asyms}),  we can set
\begin{equation}
    ( \partial_r \alpha-\partial_u \alpha + A_u )|_{u=\pm \infty}  = -{\cal Q}(r),
\end{equation}
so that,  using (\ref{eq:asym2}),
\begin{equation}
  ( \partial_r  A_u  +\partial^2_r \alpha)|_{u=\pm \infty}  =-\partial_r {\cal Q} (r).
  \label{eq:aru}
\end{equation}
Substitution of (\ref{eq:aru}) into (\ref{eq:asymh}) to eliminate $\partial_r  A_u$ gives
\begin{equation}
     4\pi r^2 J^u |_{u=\pm \infty} =-\bigg ( \partial_r (r^2 \partial_r ( \partial_r \alpha +{\cal Q}))
         +{\partial_r \alpha_{:B}}^{:B} ) \bigg )|_{u=\pm \infty} .
         \label{eq:poiss}
\end{equation}

Here (\ref{eq:poiss}) is a Poisson equation for $ ( \partial_r \alpha +{\cal Q})$.
Outside the compact support $r\le R$ of the charge-current distribution, it determines
the solution with spherical harmonic components
\begin{equation}
                  {\cal Q}(r) = \frac{Q}{r},
\end{equation}
\begin{equation}
                  \partial_r \alpha_{\ell m}|_{u=\pm \infty} = \frac {a_{\ell m} }{r^{\ell +1}}, \quad \ell\ge 1,
\end{equation}
where $Q$ is the conserved charge of the system and $a_{\ell m}$ are constants.
As a result, (\ref{eq:aru}) gives
\begin{equation}
     (r^2  \partial_r A_u) |_{{\mathcal I}^+ ,u=\pm \infty} 
    = \sum_{\ell=1, |m| \le \ell}^{\ell=\infty}\frac{(\ell+1) a_{\ell m}}{r^\ell} Y_{\ell m}  |_{{\mathcal I}^+} 
     +Q =Q.
\end{equation}
Thus
\begin{equation}
    \Delta  (r^2  \partial_r A_u) |_{{\mathcal I}^+}  = \Delta Q =0
\end{equation}
and  (\ref{eq:mem1})  governing the radiation memory reduces to
\begin{equation}
    \int_{-\infty}^{\infty}{ E_B}^{:B} du |_{{\mathcal I}^+}  = 0 .
    \label{eq:memo}
\end{equation}
Since $E_B =\partial_B(A_u -\partial_u \alpha)$ is a gradient, there are no
nontrivial solutions to (\ref{eq:memo}),
which are well-behaved on the sphere. 
As a result, the E mode radiation memory due to a bound charge-current
distribution vanishes.

\section{B mode radiation memory}
\label{sec:bm}

In order to investigate B mode radiation fields we set $A_B=\epsilon_{BC} \beta^{:C}$. 
The hypersurface equation (\ref{eq:hyp}) then reduces to
\begin{equation}
     4\pi J^u= \frac{1}{r^2} \partial_r \{r^2 \partial_r A_u ),
 \end{equation}
the dynamical equation (\ref{eq:dyn}) reduces to
\begin{equation}
     4\pi J^B= \frac{1}{r^2} \partial_r \{ 2\epsilon^{BC}\partial_u \beta_{:C}  -  {A_u}^{:B}
                    -\epsilon^{BC}   \partial_r \beta_{:C}\}
         +\frac{1}{r^4}  \{  \epsilon^{DC} { \beta_{:C}}^{:B}-\epsilon^{BC} { \beta_{:C}}^{:D} \}_{:D} 
\end{equation}
and the supplementary condition (\ref{eq:supp}) reduces to
\begin{equation}
    4\pi r^2  J^r  =  {{ A_u}^{:C}}_{:C}
                          -r^2 \partial_u \partial_r A_u.
\end{equation}

By means of the commutation relation for covariant derivatives on the unit sphere,
\begin{equation}
       v_{C:BA} -v_{C:AB} =q_{AC}v_B-q_{BC}v_A,
\label{eq:commut}
\end{equation}
the dynamical equation becomes
\begin{equation}
     4\pi J^B= \frac{1}{r^2} \partial_r \{ 2\epsilon^{BC}\partial_u \beta_{:C}  -  {A_u}^{:B}
                    -\epsilon^{BC}   \partial_r \beta_{:C}\}
         -\frac{1}{r^4}  \epsilon^{BC}{{ \beta_{:D}}^{:D}}_{:C} .
\end{equation}
The electric field satisfies $E_B = \partial_B A_u-\epsilon_{BC} \partial_u \beta^{:C}$.
The magnetic part is determined by
\begin{equation}
             \epsilon^{BC}E_{B:C} = -\partial_u {\beta^{:C}}_{:C} 
\end{equation}
so that the B mode radiation memory is determined by
\begin{equation}
          \int_{-\infty}^{\infty}   \epsilon^{BC}E_{B:C}du |_{{\mathcal I}^+} 
          = -\Delta {\beta^{:C}}_{:C} |_{{\mathcal I}^+} .
\end{equation}

The curl of the dynamical equation governs the magnetic part of the field,
\begin{equation}
     4\pi \epsilon_{BC}J^{B:C}=
      \frac{1}{r^2}\partial_r \{ 2 \partial_u{ \beta_{:C}}^{:C}   -  \partial_r  {\beta_{:C}}^{:C}  \} 
         -\frac{1}{r^4}{{{\beta_{:C}}^{:C}}_{:D}}^{:D}  .
         \label{eq:jab}
\end{equation}
We require that the Maxwell field be asymptotically stationary at early and late retarded times,
so that $\partial_u F_{\alpha\beta} \rightarrow 0$ in the limits $u\rightarrow \pm \infty$. 
For the magnetic field component $F_{AB}$ this requires that 
\begin{equation}
       \epsilon^{BC} \partial_u  A_{B:C} |_{u=\pm \infty} 
       = \partial_u  {\beta_{:C}}^{:C} |_{u=\pm \infty} =0.
\end{equation}
Thus, at early and late times, (\ref{eq:jab}) implies a Poisson equation
for  $ {\beta_{:C}}^{:C} $,
\begin{equation}
     4\pi K=  - \partial_r ^2  {\beta_{:C}}^{:C} 
         -\frac{1}{r^2}{ {{ \beta_{:C}}^{:C}}_{:D}}^{:D} , \quad u=\pm \infty ,
         \label{eq:jpois}
\end{equation}    
where  $K:=r^2\epsilon_{BC}J^{B:C}$.   
A spherical harmonic decomposition gives
\begin{equation}
     4\pi K_{\ell m}= -   \partial_r ^2  \gamma_{\ell m}
         +\frac{\ell(\ell+1)}{r^2}  \gamma_{\ell m}, \quad \ell\ge1 , \quad u=\pm \infty ,
         \label{eq:sphpois}
\end{equation}
where  $\gamma=  {\beta_{:C}}^{:C}$.
The solution is
\begin{equation}
   \gamma_{\ell m}(r) =\int_0^{\infty} r r'  g(r,r')  K_{\ell m} (r') dr'
         \label{eq:green}
\end{equation}
in terms of the Green function~\cite{jackson}
\begin{eqnarray}
           g(r,r')&=& \frac{4\pi}{\ell+1} \frac{r'^\ell}{r^{\ell+1}} , \quad r'< r \nonumber \\
           g(r,r')&=& \frac{4\pi}{\ell+1} \frac{r^\ell}{r'^{\ell+1}} , \quad r' >r .
\end{eqnarray}
Thus
\begin{equation}
   \gamma_{\ell m}(r) = \frac{4\pi}{\ell+1} \int_0^r    \frac{r'^{\ell+1}}{r^{\ell}} K_{\ell m} (r') dr'
      +\frac{4\pi}{\ell+1} \int_r^\infty  \frac{r^{\ell+1}}{r'^{\ell}} K_{\ell m} (r') dr'.
         \label{eq:greenint}
\end{equation}

The convergence of these integrals as $r\rightarrow \infty$
depends upon the limiting behavior of $K_{\ell m}(r)$. For that purpose, we model the current
by $J^\alpha = \rho u^\alpha$, in terms of a charge density $\rho$ and 4-velocity $u^\alpha$.
The momentum density is then given by $\mu u^\alpha$, in terms of the mass density $\mu$,
and the Cartesian components of the angular momentum density are
$M_i= \epsilon_{ijk}\,  \mu \, x^i u^k$.
It follows that $\mu u^i =O(r^{-4-\epsilon})$ ($\epsilon >0$) for a system with finite
angular momentum. Assuming a finite charge to mass ratio $\rho/ \mu$, this
implies that the spatial components of current satisfy $J^i =O(r^{-4-\epsilon})$.
As a result of the transformation to spherical coordinates, $J^A =O(r^{-5-\epsilon})$
and thus $K=r^2\epsilon_{BC}J^{B:C}=O(r^{-3-\epsilon)})$.

This asymptotic behavior of $K$ is sufficient to guarantee the convergence of the integrals
in (\ref{eq:greenint}). The limiting behavior is given by
\begin{equation}
   \gamma_{\ell m}|_{{\mathcal I}^+} = \frac{4\pi}{\ell+1} \lim_{r\rightarrow \infty}
   \int_0^r    \frac{r'^{\ell+1}}{r^{\ell}} K_{\ell m} (r') dr'
   =- \frac{4\pi \ell}{\ell+1} \lim_{r\rightarrow \infty}
  \frac{r^{\ell+1}}{r^{\ell-1}} K_{\ell m} (r) =0 ,
\end{equation}
where the last equality follows from l'Hospital's rule.
As a result,
\begin{equation}
       \gamma|_{{\mathcal I}^+} =  {\beta_{:C}}^{:C}|_{{\mathcal I}^+} =0, \quad u=\pm \infty ,
\end{equation}   
and $\beta|_{{\mathcal I}^+} =0$, $u=\pm \infty$, is the only non-trivial solution.    
Thus $\Delta \beta =0$ and there cannot be any B mode radiation memory
due to a physically realistic charge-current distribution.

\section{Conclusion}

We have established two global properties of electromagnetic radiation memory:
(i) production of E mode radiation memory requires an unbound system of charges;
(ii) B mode radiation memory cannot be produced by any physically realistic charge-current
distribution. The methods used here pave the way for a similar analysis in linearized gravitational
theory, where the analogous global results for gravitational radiation memory are plausible. 

The global properties of the electromagnetic and  gravitational
radiation memory in the fully nonlinear Einstein-Maxwell theory can also be tackled by means
of a coupled characteristic initial value problem.  
The characteristic treatment of gravitational radiation has led to a robust and accurate
computational tool for extracting gravitational waveforms from
strong sources~\cite{winrev}.  The coupling to the electromagnetic field leads to
to  gravitational radiation memory due to the flux of electromagnetic radiation
to null infinity~\cite{zipser,yau1,yau2}, analogous to the flux of gravitational energy which produces
the nonlinear memory discovered by Christodoulou. These forms of radiation memory
have been referred to as ``null memory''~\cite{bieri2}, to clarify that they do not necessarily
require nonlinearity but only a flux of energy to null infinity. 
This also includes zero rest mass particles~\cite{wald,tolish} and neutrinos~\cite{neutmem}.
Such null memory arises
in the same way as nonlinear gravitational memory and is of the electric type
(see Appendix~\ref{sec:app}).

In the context of the BICEP2 experiment, the interaction with gravitational waves has the crucial
effect of producing B mode electromagnetic radiation. The extent to which relativistic gravitational
effects, including black hole formation, can lead to B mode electromagnetic radiation
memory is an interesting question. Computational simulation might shed light  on this issue..

In addition to the possibility of observing gravitational radiation memory using pulsar timing
arrays~\cite{pulsta}, there are also compelling theoretical concerns that warrant further
study of the memory effect.
Gravitational radiation memory is related to the supertranslation freedom in the 
Bondi-Metzner-Sachs (BMS) group~\cite{bms}, which
is the asymptotic symmetry group for asymptotically flat space times.
For a stationary system, this supertranslation freedom can be removed and the BMS group
reduced to the Poincare group, in which case energy-momentum
and angular momentum are well-defined.
However, for a system which makes a stationary to stationary transition, the two Poincare
groups obtained at early and late times can differ by a supertranslation. This in fact occurs
if the intervening gravitational radiation produces a non-zero memory.
See~\cite{newpbms,gw,helfer} for discussions. (These considerations were already presaged
in Part D of the pioneering work of Bondi~\cite{bondi}.)
Such supertranslation shifts of the electric type could lead to a
distinctly  general relativistic mechanism for a system
to lose angular momentum.  A non-zero magnetic type supertranslation shift would
have a more drastic effect on a proper understanding of angular momentum
in general relativity.

\begin{acknowledgments}

This work was motivated by the seminal results of  L. Bieri and D. Garfinkle,
which I learned of as guest of the Mathematical Sciences Research Institute,
supported by NSF grant 0932078 to the University of California at Berkeley.
My research was supported by NSF grant PHY-1201276 to the University of Pittsburgh.

\end{acknowledgments}

\appendix

\section{Global decomposition of gravitational radiation memory}
\label{sec:app}

Gravitational radiation has a decomposition analogous to the E and B
mode decomposition of electromagnetic radiation. We use the notation of Sec.~\ref{sec:emrm}.
The strain tensor which measures the gravitational radiation
is normally represented as a trace free tensor field $\sigma_{AB}(u,x^C)$ on the sphere.
Equivalently, in terms of a polarization dyad it can be represented as a spin-weight
2 field $ \sigma=q^A q^B h_{AB}$, The decomposition into electric and magnetic
electric and magnetic parts is achieved by relating $\sigma$ to a spin-weight-0
field $\Sigma$ by means of the spin-weight raising operator $\eth$~\cite{goldberg},
\begin{equation}
      \sigma= \eth^2  \Sigma := q^A q^B\Sigma_{:AB}.
\end{equation}
The decomposition of  the spin-weight-0 potential $\Sigma$ into real and imaginary parts, 
\begin{equation}
   \Sigma(u,x^A) =\alpha(u,x^A)+i\beta(u,x^A),
\end{equation}
is the gravitational analogue of the E and B mode decomposition. The electric part 
$\alpha$ has even parity and the magnetic part $\beta$ has odd parity.
The corresponding modes of gravitational wave memory are determined  by
the real and imaginary parts of
\begin{equation}
      \Delta \Sigma = \Delta (\alpha+i \beta) =(\alpha+i\beta)|_{u=\infty}-(\alpha+i\beta)|_{u=-\infty}.
\end{equation}

It is easy to show, analogous to the electromagnetic examples presented in Sec.~\ref{sec:emrm},
that both electric and magnetic type gravitational radiation memory can be realized in linearized
theory by source free gravitational waves.
It is the purpose of this Appendix to show that the two known sources of
gravitational memory based upon physically realistic systems, the
``linear'' and ``nonlinear'' memory,  both produce memory of electric type. 

First consider the ``linear'' radiation memory produced by the ejection of a massive particle with
velocity $V_i$ from an initially bound system. In the present notation,  the  results of
Braginsky and Thorne~\cite{brag} lead to the radiation memory
\begin{equation}
         \Delta \sigma = \frac {4M \, Q^i V_i \,  Q^j V_j}{\sqrt{1-V^2} \, (1-V_i n^i)},
         \label{eq:brag}
\end{equation}
where $Q^i =\eth n^i =q^A {n^i}_{:A}$ is a complex basis for the 2-space transverse
to the radial unit vector $n^i$.
Because $n^i=(\sin\theta \cos\phi, \sin\theta\sin\phi,\cos\theta)$ consists of $\ell=1$
spherical harmonics, it follows that $\eth^2 n^i=0$, i.e. spin-weight-2 functions are
composed of spherical harmonics with $\ell\ge 2$. As a result, we can
rewrite (\ref{eq:brag})  in terms of a real spin-weight-0 potential as
\begin{equation}
         \Delta \sigma = \frac {4M }{\sqrt{1-V^2}} \eth^2 \bigg ( (1-n^i V_i)\ln (1-n^i V_i) \bigg ).
\end{equation}
Thus the ``linear'' memory is purely of
the electric type.

Next consider the  ``nonlinear'' memory discovered by Christodoulou.
A simple formulation has been given by Frauendiener~\cite{frauen} in terms of the
Newman-Penrose formalism~\cite{np}, in which the radiation field $\sigma$
is the asymptotic shear of the outgoing null hypersurfaces $u=const$.
He finds that the nonlinear memory satisfies
\begin{equation}
    \eth^2  \Delta \bar \sigma = -\Delta( \Psi_2 +\sigma \bar {\dot \sigma}) 
            +\int_{-\infty}^{\infty} |\dot \sigma |^2 du,
            \label{eq:gmem}
\end{equation}
where $\Psi_2$ is the leading asymptotic part of the Weyl tensor component.

The last term in (\ref{eq:gmem}) is the nonlinear memory produced by the
radiation of gravitational energy to ${\mathcal I}^+$.
It is manifestly real and leads to
radiation memory of the electric type.

The first term in (\ref{eq:gmem}) incorporates the radiation
memory produced by freely escaping massive particles
in linearized theory, which is again of the electric type.
It remains an open question whether this
first term can lead to magnetic type radiation memory.
One way to check  whether it has a magnetic component is to
calculate
\begin{equation}
      \bar \eth^2   \Delta \sigma -  \eth^2 \Delta \bar \sigma 
      =  \bar \eth^2 \eth^2 (\alpha+i \beta) -  \bar \eth^2 \eth^2 (\alpha- i\beta)
      = 2i  ( \bar \eth \eth)^2  \beta,
      \label{eq:ethbeta}
\end{equation}
where the last equality follows from  the general commutation relation 
\begin{equation}
         (\bar \eth \eth - \eth \bar \eth) F = 2s F
\end{equation}
for a spin-weight-$s$ quantity $F$. (This commutation relation is the $\eth$ version of
the commutator of covariant derivatives on the unit sphere (\ref{eq:commut}).)
Applied to (\ref{eq:gmem}), this gives 
\begin{equation}
      \bar \eth^2 \Delta  \sigma -  \eth^2 \Delta \bar \sigma
      = \Delta \bigg( ( \bar \Psi_2 +\bar \sigma {\dot \sigma}) 
            -  ( \Psi_2 +\sigma \bar {\dot \sigma}) ) \bigg ).
            \label{eq:ethmem}
\end{equation}
But in the Newman-Penrose formalism (see~\cite{frauen})
\begin{equation}
   \Psi_2 -\bar \Psi_2 = \bar \eth^2 \sigma -\eth^2 \bar \sigma 
   +\bar \sigma \dot \sigma - \sigma \bar {\dot \sigma} .
\end{equation}
As a result, (\ref{eq:ethmem}) reduces to the identity
\begin{equation}
      \bar \eth^2 \Delta  \sigma -  \eth^2 \Delta \bar \sigma 
      = \Delta ( \bar \eth^2 \bar \sigma    -  \eth^2 \bar  \sigma ).
\end{equation}
So it appears that this type of purely asymptotic argument cannot be
used to study gravitational radiation memory of the magnetic type.

\end{document}